# The University AI Didn't Replace: Rethinking Universities in the AI Era

Karol P. Binkowski and Andrew Hopkins, Macquarie University

karol.binkowski@mq.edu.au, andrew.hopkins@mq.edu.au

## Abstract

Generative artificial intelligence (AI) is reshaping higher education, yet many universities remain in early stages of adoption where AI innovation occurs informally and without institutional recognition. This paper presents a framework describing four levels of AI adoption in universities and illustrates these dynamics through a case study of AI-enabled curriculum initiatives in several units. We contend that the key institutional challenge is moving from isolated innovation to strategic integration, where universities redesign learning around AI-supported reasoning and align policies, workload models, and recognition systems to support educational transformation.

## Introduction

The rapid emergence of generative artificial intelligence (AI) is reshaping debates about the role of universities in a knowledge-rich digital environment. Large language models capable of generating explanations, analyses, and written outputs challenge traditional approaches to teaching and assessment, raising questions about how universities cultivate higher-order thinking and disciplinary reasoning (Kasneci et al., 2023; Mollick & Mollick, 2024). International policy bodies and assessment scholars increasingly argue that responding to generative AI requires systematic redesign of curriculum and assessment practices alongside institutional support for educators experimenting with AI-enabled pedagogy (Bearman et al., 2023; UNESCO, 2023). Commentators have also noted that educators often lack sufficient institutional time and recognition to explore AI-driven innovations in teaching and assessment (Liu, 2026). Recent sector discussions suggest that generative AI is less a disruptive force and more a mirror reflecting long-standing tensions in higher education, particularly around what constitutes learning and how it should be assessed (Lazarus et al., 2026). The central challenge for universities is therefore not how students will use AI, but how institutions redesign learning environments so that AI supports the development of reasoning, critical evaluation, and responsible professional practice. This aligns with emerging national calls to reconceptualise learning and assessment toward deep understanding and capability rather than the production of outputs (Castlereagh Summit, 2026).

As AI increasingly performs technical and analytical tasks, the purpose of education shifts from producing answers to developing human capabilities in reasoning, judgment, and the ability to act under uncertainty. This aligns with a people-centred view of education, where students are the primary outcome of the learning process, not merely a means to outputs. If AI is allowed to perform the core intellectual work, students risk disengaging from the process of becoming practitioners. This highlights the need to redesign curriculum and assessment around active reasoning and reinforces that such transformation requires explicit workload recognition and institutional support for educators (Hogg, 2026).

This shift also reflects a broader recognition that intelligence is not a single continuum on which humans are simply outperformed, but a set of distinct capabilities shaped by human constraints, including learning from limited experience, creativity, and meaning-making (Griffiths, n.d.)

**Time and effort**

Developing and implementing these innovations has required substantial time, experimentation, and pedagogical redesign. However, such efforts are often insufficiently recognised within institutional structures, including workload models, promotion criteria, and teaching awards. More broadly, as universities move forward, leadership often draws on examples from other institutions when proposing change rather than inward to the work of local innovators when identifying solutions to emerging educational challenges. This dynamic is not unique to one institution but reflects a wider pattern across the sector, where innovation in AI-enabled education risks stagnation because the individuals experimenting with new approaches operate without structural recognition or strategic support. This reflects broader sector-wide observations that meaningful transformation is constrained by fragmented institutional responses and insufficient alignment of workload, incentives, and support structures (Castlereagh Summit, 2026).

This paper is motivated by these observations and calls for policy changes that recognise and support AI-driven curriculum innovation within universities.

**Levels of AI Adoption in Universities**

Universities currently vary in their responses to generative AI in teaching and assessment. The following four-level framework summarises different stages of institutional response, ranging from defensive restriction to full integration of AI into educational systems, with many institutions presently operating around Level 0 or Level 1.

**Level 0 — Defensive Containment**

At this level, AI is treated primarily as a threat to academic integrity. Institutional responses focus on restricting AI use rather than adapting teaching practices.

- AI discouraged or banned in assessment
- Heavy reliance on invigilated exams
- AI framed mainly as an integrity risk
- Little or no curriculum redesign

This approach often creates a mismatch between policy and the reality of widespread AI use by students.

**Level 1 — Informal or Peripheral Adoption**

Here, experimentation with AI occurs largely at the level of individual educators rather than through institutional strategy. Innovation is driven by local initiatives.

- Individual lecturers experiment with AI-enabled tasks
- Limited or inconsistent guidance for students
- No formal recognition in workload or promotion
- Uneven adoption across units and programs

As a result, innovation depends on motivated individuals and may place additional pressure on those leading change.

**Level 2 — Strategic Integration**

At this stage, universities begin integrating AI into teaching policy and curriculum design. Institutional structures start to support innovation.

- Clear institutional principles on AI use
- Curriculum redesign acknowledging AI-rich environments
- Professional AI development for educators
- Recognition of AI-related innovation in workload or funding

Teaching increasingly focuses on reasoning, interpretation, and responsible AI use.

**Level 3 — AI-Embedded or Transformational Universities**

In this stage, AI becomes part of the core educational infrastructure and learning ecosystem.

- AI integrated into core university infrastructure (LMS, assessment, learning support)
- Curriculum centred on AI-supported reasoning rather than content delivery
- Assessments evaluate judgment, justification, and decision-making in AI-rich environments
- Students routinely work with AI as a cognitive partner in authentic problem solving
- Industry-integrated learning reflecting real-world AI-enabled professional practice

Universities at this level position AI as a foundational component of teaching while focusing on human reasoning and decision-making skills.

The **bottom-left quadrant (Level 0 – Prohibition)** on Figure 1 represents institutions where AI is largely restricted, and teaching practices remain traditional. The **bottom-right quadrant (Level 1 – Informal Innovation)** reflects environments where individual educators experiment with AI, but without institutional recognition or coordinated strategy. The **top-left quadrant (Controlled Adoption)** represents institutions where AI is acknowledged in policy discussions but has not yet significantly transformed teaching practice. Finally, the **top-right quadrant (Level 2+ – Strategic Integration)** represents universities where AI is systematically embedded in curriculum, assessment, and institutional structures supporting educational innovation.

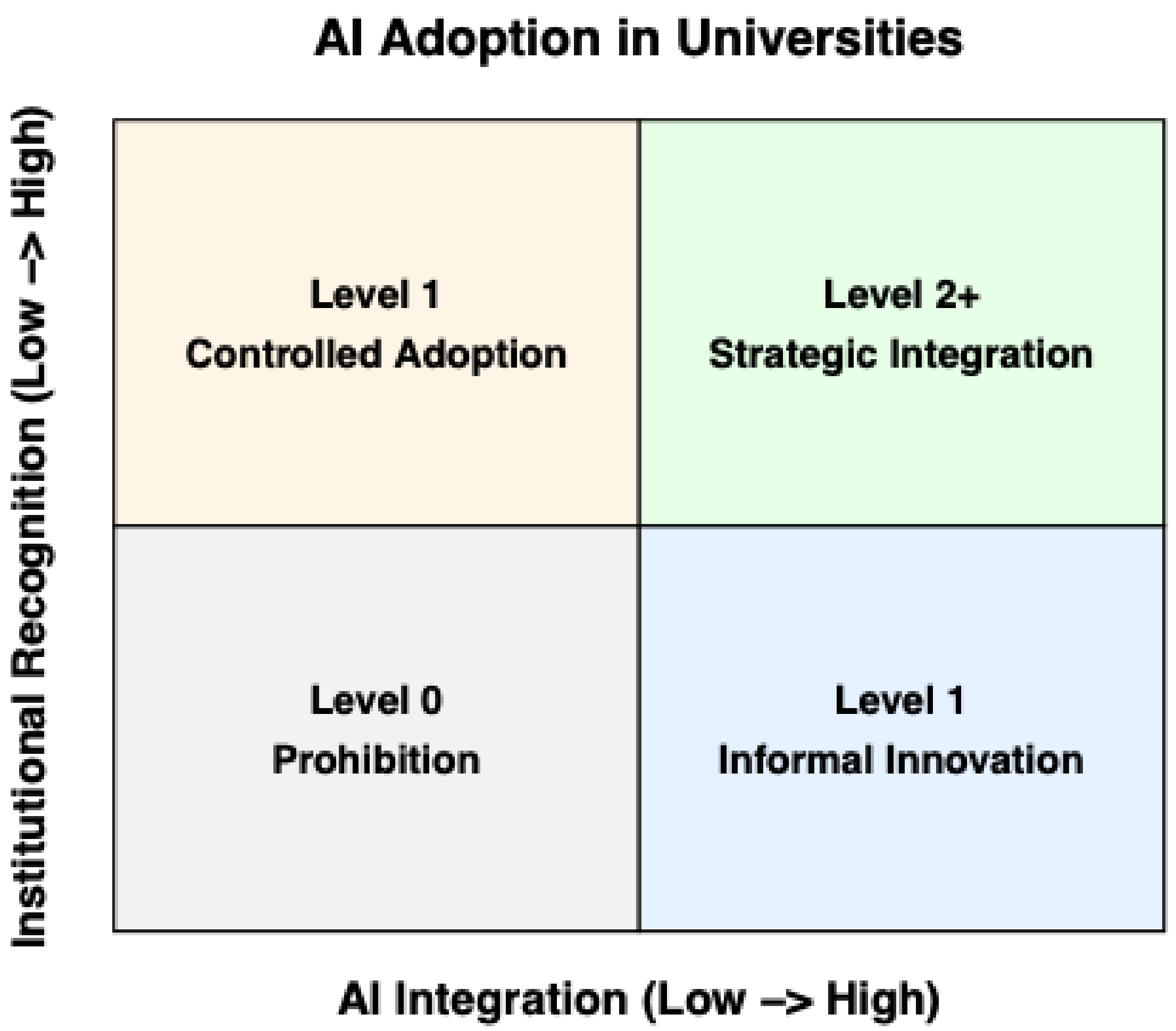

Figure 1. A **2×2 matrix describing levels of AI adoption in universities**. The horizontal axis represents the degree of **AI integration in teaching and curriculum**, while the vertical axis represents **institutional recognition and structural alignment**, including policies, workload models, and promotion frameworks.

**Learning Across Levels of AI Adoption in Universities**

Learning in Levels 0–1 largely resembles traditional university models, centred on content delivery and assessments that either assume the absence of AI or allow limited, informal experimentation introduced by individual educators. At Level 2, learning begins to systematically integrate AI into curriculum design, with students developing AI literacy and learning how to use AI tools critically alongside conceptual understanding and disciplinary reasoning. At Level 3, learning is fundamentally reorganised around AI-supported reasoning, with students routinely working with AI as a cognitive partner while demonstrating judgment, justification, and problem-solving in authentic, often industry-connected contexts.

**Case Studies**

**AI-Aware Teaching in a Large Introductory Statistics Course.** This paper draws on experience of implementing innovations in STAT1170, a large introductory statistics unit at Macquarie University enrolling approximately 1200 undergraduate students per session from diverse non-statistical disciplines. The AI-related innovations in the unit build on an established teaching approach designed to engage large, diverse cohorts in learning statistics. Based on Peer Instruction sequences embedded within live lectures, conceptual discussions aimed at revealing misconceptions, and created with the use of genAI narrative-based learning through the storyline of *Detective Sigma* and *Dr Means*. Building on this foundation, several AI-enabled elements were introduced, including AI-assisted statistical poster assessment, AI-enhanced pre-recorded lectures, and a university-provided AI "Virtual Peer" a chatbot tool that offers students on-demand study support based on unit materials. Together, these interventions aim to support engagement and conceptual understanding while positioning AI as a tool that complements, rather than replaces, statistical reasoning.

Beyond **STAT1170**, a broader set of case studies across disciplines illustrates how the university is beginning to respond to generative AI through assessment redesign rather than restriction. These examples can be grouped into three emerging approaches: AI-supported independent work, process-oriented and reflective assessment, and observed or authentic performance-based tasks.

The first group includes **AI-supported independent assessments**, where AI use is explicitly permitted but bounded by expectations of student accountability. Examples include the *Protein Portfolio* in **BMOL2201/6201**, where students iteratively develop scientific understanding over six weeks with AI supporting feedback and referencing, and the *GenAI-assisted statistical poster* in **STAT1170**, where students analyse individual datasets and may use AI for visualisation or structuring outputs. In both cases, AI is integrated into the workflow, but students remain responsible for interpretation, synthesis, and disciplinary reasoning.

The second group reflects a shift toward **process-oriented and reflective assessment**, where learning is demonstrated through development over time rather than final outputs. In **APPL8241**, students complete weekly seminar tasks and later analyse and reflect on their work, while in **INTS1001**, a combination of portfolio, reflective writing, and viva voce captures students' engagement with systems thinking across multiple weeks. These designs emphasise metacognition and reflection, with AI positioned as a support tool rather than a substitute for thinking.

The third group represents **observed and authentic assessment**, where student understanding is evaluated through live performance and interaction. In **MKTG1001**, students deliver a *Shark Tank* pitch supported by a poster and respond to questioning to demonstrate individual understanding, while in **MMCC2141**, students present an editorial pitch grounded in media theory. These tasks require students to justify, explain, and adapt their thinking in real time, aligning more closely with professional practice in AI-rich environments.

Doosti (2026) provides a further case study of AI-enabled curriculum innovation at Macquarie University, illustrating a Level 1-to-2 transition in a specific disciplinary context. Delivered to over 250 undergraduate students at Nanjing Normal University under the NNU-MQ Joint Institute, the unit STAJ1371 Statistical Data Analysis embedded structured AI use for practice generation, code explanation, and scalable formative feedback, while deliberately centring statistical thinking rather than procedural execution as the core learning goal. The case illustrates how an individual educator, operating with pedagogical intent and a coherent clinician-inspired framework, can move AI integration beyond informal experimentation toward the kind of reasoning-focused, assessment-redesigned practice that requires institutional recognition to scale.

London School of Innovation (LSI) presents itself with a strong "AI-native university" brand, positioning higher education as needing fundamental redesign rather than incremental adaptation to generative AI, while Singapore University of Technology and Design (SUTD) approaches AI within a broader interdisciplinary and design-centred educational philosophy. Both institutions are experimenting with AI-enabled personalised learning and a shift toward more human-centred teaching roles focused on mentoring, collaboration, judgement, and problem-solving rather than on traditional content delivery alone. While LSI currently appears more vision- and branding-driven and SUTD more institutionally embedded and operationalised, together they illustrate emerging models of universities attempting to structurally integrate AI into curriculum, pedagogy, and institutional design. Such institutions are already positioning themselves as AI-enabled education enterprises (https://lsi.ac.uk/, https://www.sutd.edu.sg/) although it is not completely clear whether these are genuinely at the "strategic adoption" stage.

Across these cases, a consistent pattern emerges: effective integration of AI requires a shift toward assessment designs that foreground reasoning, justification, and process over product. However, such a redesign is significantly more complex and time-intensive than traditional assessment, reinforcing the need for institutional structures that recognise and support this work.

In the Australian context, the accreditation body TEQSA (Tertiary Education Quality and Standards Agency) has been developing guidelines to assist universities in accommodating AI tools within their assessment structures (Tertiary Education Quality and Standards Agency, 2026). Again, it remains open as to how and whether universities adopting these recommendations will do so in a high-level and strategic fashion (Level 2), or in a more reactionary, ad hoc, or fragmented way (Level 1).

**Risks of Remaining in Early Stages of AI Adoption**

Universities that fail to progress beyond Levels 0 or 1 to reach Level 2 risk falling into a reactive mode where policies attempt to contain AI rather than harness it for learning, leading to fragmented teaching practices, growing misalignment between assessment and real-world practice, and missed opportunities to prepare students for AI-enabled professional environments.

If universities do not move toward Level 3, they risk losing economic relevance as governments and industry partners increasingly direct funding toward institutions that demonstrate leadership in AI-enabled education. However, a sector-wide concern is that universities frequently look to peer institutions before making major changes, creating a tendency toward cautious imitation that may significantly slow the pace of necessary transformation.

These challenges are compounded by infrastructural and cultural barriers. As Kezar (2018) argues, many change efforts falter because leaders and educators operate with implicit and overly simplistic theories of change, assuming that tools or individual experimentation will be sufficient. Institutional systems remain geared toward traditional practices and offer little support for AI-enabled redesign. Without deliberately building shared sensemaking and a culture of transformation, individual innovations typically remain isolated, unrecognised in workload and promotion structures, and unable to scale into strategic institutional change.

Key strategic steps to move beyond Level 1 of informal innovation:

- Recognise AI-driven curriculum redesign in workload models so innovation is not dependent on unrecognised effort by individual educators.
- Embed AI-enabled teaching innovation in promotion and teaching award criteria, signalling institutional value and legitimacy.
- Establish clear institutional principles for AI use in learning and assessment to move beyond fragmented, unit-level experimentation.
- Redesign assessment toward reasoning and justification in AI-rich environments, rather than attempting to merely tolerate AI.
- Create institutional pilots or funded initiatives that scale successful teaching innovations beyond individual courses.

**Conclusion and discussion**:
Ultimately, the path to meaningful progress in the AI era does not solely involve distributing more AI tools to students and educators but also depends on strategically dedicating time and institutional support to enable educators to thoughtfully redesign curricula and assessments. By emphasising workload recognition and structural incentives for pedagogical innovation rather than just technological provision, universities can speed up the shift from informal experimentation to transformative, AI-supported learning environments that genuinely equip students for the future.

The arrival of generative artificial intelligence represents one of the most significant technological shifts affecting higher education in decades. While many universities have begun experimenting with AI-enabled teaching and learning, most innovation remains fragmented and dependent on individual initiative. The central challenge is not technological adoption but institutional alignment. Universities must redesign curricula and assessment while recognising and supporting the educators who lead these transformations. By moving from informal experimentation toward strategic integration, universities can create learning environments in which AI enhances rather than undermines the development of human expertise.